# A route towards controlling the morphology of vertical graphene nanosheets

Subrata Ghosh* [1,2], K. Ganesan[2], S.R. Polaki[1,2] and M. Kamruddin[1,2]

[1]Homi Bhaba National Institute, Mumbai – 400094, India

[2]Surface and Nanoscience Division, Materials Science Group, Indira Gandhi Centre for Atomic Research, Kalpakkam - 603102, India

**Abstract**

Herein, an effort has given to sheds light on the effects of plasma process parameters on the growth of vertical graphene nanosheets (VGNs) by plasma enhanced chemical vapor deposition (PECVD. The parameters include substrate temperature, microwave power and distance between plasma sources to substrate. The significant influence of these variable parameters is observed on the morphology, growth rate and crystallinity. Thus these parameters are found to be deciding factors, which determine the surface reaction and growth kinetics that governed the final structure and controlled morphology of VGNs. The activation energy of the VGNs grown by PECVD is found to be 0.57 eV. A direct evidence of vertical growth through the nanographitic island is observed from temperature dependent growth of VGNs. Such understanding on growth of VGNs is not only useful for growth mechanism under plasma chemistry but also beneficial to get controlled and desired structure for field emission and energy storage application.

\* subrataghosh.phys@gmail.com (Subrata Ghosh)

**Introduction**

The vertical graphene nanosheet (VGNs), one of the close relatives of graphene, has attracted current research interest due to its unique geometry and remarkable properties. It is an interconnected porous network of vertical sheets, also known as carbon/graphene nanowalls/nanosheets. Each sheet of VGNs is stacked of 5-12 graphene layers. The height and length of each sheet can be varying upto few micron. The VGNs feature huge surface area, sharp edges, good electrical conductivity and adaptive capability of functionalization. Therefore, this novel nanoarchitecture already received envisaged opportunity for next generation electron emission, sensors, energy storage nanodevices and bio application.[1]

The VGNs are extensively grown by PECVD technique, which allows growing at low temperature by means of plasma.[1-4] Even different kind of novel plasma processing unit has designed extensively and well addressed to understand nucleation and growth of VGNs by Hiramatsu group.[4] Several groups have discussed the effect of various process parameters e.g. career gas, nature of substrates, total pressure, microwave power on the growth of VGNs.[5-28] The underlying mystery behind the growth mechanism of VGNs under plasma atmosphere still remains unresolved.[1] The structure and morphology of PECVD-grown nanostructure is strongly determined by substrate-plasma species interaction. The substrate defines the surface reaction kinetics, thus nature of substrates and substrate temperature plays major role on the growth. In Plasma based growth of carbon nanostructure, the hydrocarbon and precursor gas under the plasma is decomposes into various ranges of plasma radicals. The amount and nature of plasma species in terms of density and energy of ions depends on the plasma power, the position of the substrate in the plasma plume and feedstock gas composition ratio at a constant substrate temperature. Therefore these key factors define the surface reaction kinetics and growth dynamics. Therefore, in order to get quite uniform, high quality, precise orientation VGNs, it is necessary to get clear image on the plasma characteristics or behavior of plasma active species under various process parameters during growth. The roles of carrier gas with hydrocarbon gas, nature of substrate and deposition time on the growth of VGNs and a phenomenological four-stage model are addressed in our previous findings.[5, 26, 29] Thus role of substrate temperature, microwave power and distance from substrates to plasma source on the growth are the concern of this study.

In this manuscript, we have grown VGNs by electron cyclotron resonance (ECR)-PECVD by varying the plasma process parameters by keeping fixed $CH_4/Ar$ gas flow and growth time. The variable parameters are growth temperature, distance between substrate to plasma source and plasma power. The systematic investigations are carried out to identify the optimal conditions for growing VGNs under PECVD with the help of scanning electron microscopy (SEM) and Raman spectroscopy.

## 2. Experimental methods

### 2.1 Growth of VGNs by ECR-CVD

The VGN films are deposited on $SiO_2$/Si substrates. The $SiO_2$ film with a thickness of 400 nm grown by thermal oxidation on n-Si(100) substrate. The details of the ECR-CVD deposition system are reported elsewhere.[5] Here we choose ultrahigh pure $CH_4$ as hydrocarbon source and commercial grade Ar as career gas. Prior to the growth, the substrates are cleaned by acetone, Isopropyl alcohol, di-ionized water followed by drying with $N_2$ gas and loaded to the chamber, separately and respectively. The chamber is evacuated to $10^{-6}$ mbar by rotary pump and followed by turbo molecular pump. Then substrates are cleaned by Ar plasma with the help of microwave (2.45 GHz) power of 200W at respective temperatures. The growth details can be found elsewhere. The deposition parameters for VGNs growth are listed in Table 1. Followed to the growth, the substrates are annealed for 30 min at respected temperature by putting off the microwave plasma. Finally, deposition unit is bringing down to room temperature for further characterization.

To investigate the growth process of VGNs under plasma, we kept the flow rates of CH4/Ar, growth time, annealing condition (before and after growth) as a constant for this study, as mentioned earlier. Only variable parameters are growth temperature, microwave power and distance between plasma source and distance. The work is classified into three cases, as shown in table 1. For simplicity, the distance between plasma source and substrate is mentioned by only distance throughout the manuscript.

### 2.2 Characterization

Morphological features of the films are observed by SEM (Supra 55, Zeiss). The wettable properties of the film are measured by sessile drop method with the help of a CCD camera (Apex Instrument Co. Pvt. Ltd., India). The volume of the droplet is about 1μl and all measurements are carried out in ambient conditions. The value of contact angle is evaluated by half angle fitting method provided with the instrument. Raman scattering measurement is carried out with a micro-Raman spectroscopy (Renishaw inVia, UK) to evaluate the structural properties in terms of defects and disorder. In backscattering geometry, spectra were taken in the 1000-3500 cm$^{-1}$ frequency region for 30s accumulation time using 514 nm laser and 100× objective lens.

Table 1: Deposition parameters for VGNs growth by ECRCVD

| | |
|---|---|
| CH$_4$/Ar gas flow (sccm) | 5/25 |
| Gas pressure (mbar) | $2.8 \times 10^{-3}$ |
| Growth time (min) | 30 |
| Case I | |
| Growth temperature (T$^o$C) at d=30 cm and MW power, P=375W | 600, 625,650 725, 800 |
| Case III | |
| Microwave power (P in Watt) at d=30 cm and T=800$^o$C | 200, 280, 320, 375,425, 475 |
| Case I | |
| Distance (d in cm) at P= 375W and T= 800$^o$C | 15, 20, 30, 40 |
| Annealing time (min) at respective growth temperature | 30 |

## 3. Result and discussion

*Case I: Influence of growth temperature at consant microwave power and distance*

We investigate the early stage nucleation and growth of VGNs by changing the substrate temperature from 600 to 800$^o$C under CH$_4$/Ar gas environment for 30 min, while plasma power and distance are fixed at 375W and 30 cm respectively. Fig. 1(a-f) represents the morphology of the films grown under various substrate temperature ranges. It is seen that an increase in substrate temperature leads to grow VGNs from the nanographitic (NG) layer. No NG structure is grown below 600$^{\circ}$C in this study as etching of graphene by hydrogen species dominates.[25] It can be seen from Fig. 1(a) that the continuous NG island can be grown under 600$^{\circ}$C and in inset of Fig.1(a) represents the higher magnified image of VGNs. The height of NG film of 17.72 (±1.64) nm, average domain size of 28.62(±17.41) nm and domain density of 37.0(±6.5) are found for 0.2 μm × 0.2 μm area. The films are found to be electrically continuous and sheet resistance of 5.56 KΩ/□. Such growth directly on insulating substrate at low temperature without post-growth treatment offers a good compatibility with the semiconductor technological. As temperature increases to 625$^{\circ}$C the two dimensional structure transformed to three dimensional structures by growth of vertical sheet on the NG layer. At 625$^{\circ}$C, the grain boundary of nanoisland domain acts as a nucleation sites for vertical growth with the help of in-built localized electric field due to plasma and thermophoretic force induced by a temperature gradient in the plasma, is shown in Fig 1(b).

As shown in Fig. 1(c), the height of vertical sheets of 37.45(±8.94) nm is observed on the NG layer of thickness of 11.69(±1.97) nm. In addition, the average domain size and domain density on the NG layer are found to 40.41(±18.26) nm and 30.75 (±2.75), respectively for 0.2 μm ×0.2 μm area. Further increase of temperature to 650°C [Fig. 1(d)] led to more number density of vertical sheet as compared to previous sample and height of and forms a nest-like structure with the higher growth rate, as shown in Fig. 1(e). Further increase in temperature, the lateral dimension of the sheets increases and the sheets started to interlace together as a result the spacing between the two vertical sheets while maintaining the sheet like features at 800°C [Fig. 1(f)]. Fig. 1(g) shows the variation of growth rate with respect to the growth temperature. It has seen that the variation of growth rate vs growth temperature follow the Arhenius equation and hence activation energy is calculated. The activation energy of growth is found to be 0.57 eV.

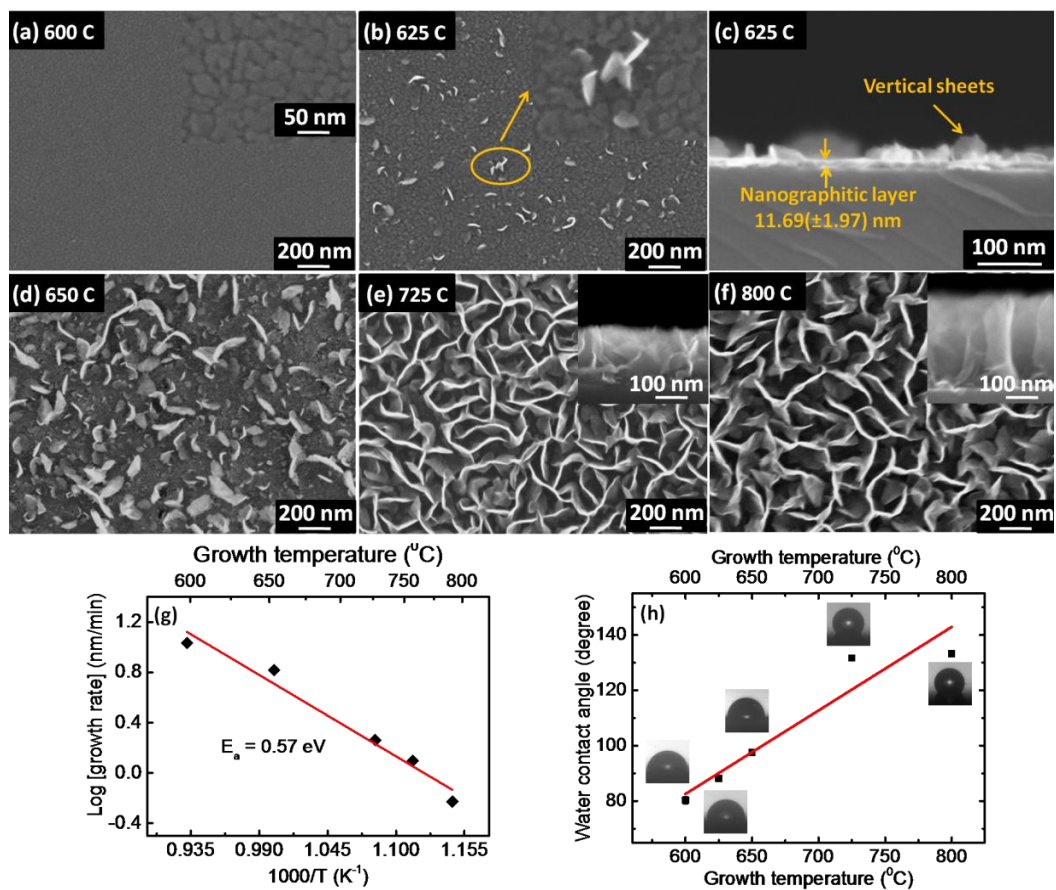

Figure 3.11 : SEM micrograph of the sample grown at (a) 600°C, (b) 625°C, (c) 625°C (cross section image), (d) 650°C, (e) 725°C and (f) 800°C, while deposition time, microwave power and distance between plasma source to substrates are 30 min, 375W and 30 cm respectively. (a) Arhenius plot of the growth rate *versus* inverse substrate temperature for case 1, (b) Water contact angle versus growth temperature of VGNs.

The CA of the surface at ambient environment with respect to the growth temperature is shown in the Fig. 1(h). The CA measurement shows that the depending on the morphology, hydrophilic surface with contact angle 80.3$^o$ (sample grown at 600$^o$C) transforms into the near superhydrophobic with contact angle 131.5$^o$ (sample grown at 800$^o$C).

Fig. 2(a) represents the Raman spectra of plasma grown samples. The first order Raman spectra of VGNs mainly consists of D peak at 1350 cm$^{-1}$, G peak at 1580 cm$^{-1}$ and D′ peak at 1620 cm$^{-1}$. Whereas, the second order Raman spectra consists of D+D″, G′, D+D′ and 2D′ bands. The defect related bands (D, D′, D″ and its overtones) confirm the presence of high edge density, structural defect and disorder, whilst the presence of G bands assure the graphitic nature in the film. The occurrence of distinct G′ bands in Raman spectra confirms that presence of graphene structure in the sample. The strong D band and other defect related peaks confirmed the huge amounts of defects are present in the structure. The defects may include vacancies and strained hexagonal/non-hexagonal (pentagon or heptagon) distortions that lead to the non-uniformity, corrugation and twisting as shown in SEM micrographs.[22] It is observed that the D′- band (G′-band) intensity increases (decreases) and started to decrease (increase) from as growth temperature increases from 625°C to 800°C [Fig. 2(a)]. Similarly result observed for the FWHM of D, G and G′-band, position of G-band and $I_D/I_G$ that the all these parameters increases as growth temperature increases to 625°C and decreases with increasing temperature from 625°C to 800°C [Fig. 2(b) and (c)]. This result suggests the transformation of structure and found to good agreement with SEM micrograph observation [Fig. 1]. At 625°C the edges of vertical sheets started to form and hence increased in defect in the structure, whereas further increase in temperature resulted with better degree of graphitization in the structure. In addition, the $I_D/I_G$ also found minimum for the sample grown at 800°C.

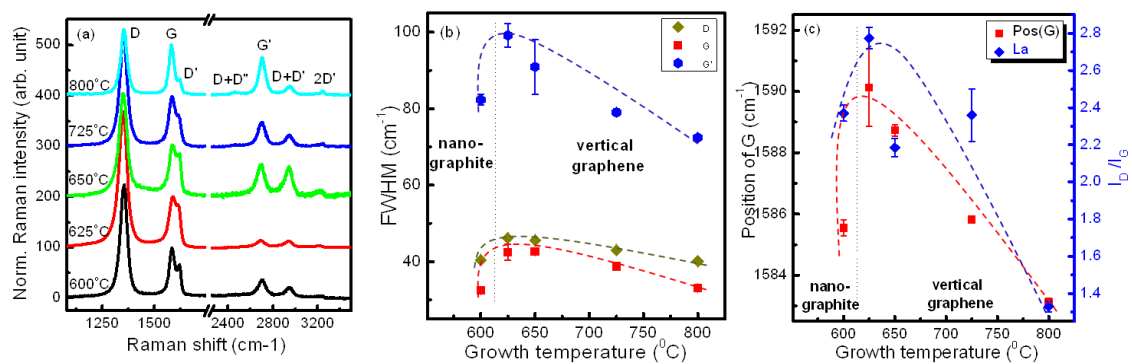

Figure 2: (a) Raman spectra of temperature dependent growth of VGNs,(b) Full width at half maxima (FWHM) of D, G, G') and $I_D/I_G$ vs growth temperature.

This observation concludes that the higher growth temperature provides higher deposition rate with better crystalline structure. Thronton et al observed that the plasma density and composition has not significant affected by substrate temperature.[30] As growth temperature increases, the mean free path of plasma species also increases as it is directly proportional. Hence, this result is only due to the increased mobility of surface atoms such that adsorption rate and chemical surface reaction becomes higher at high temperature which minimizes the surface energy of the substrate. During pre-cleaning procedure, the Ar plasma removes surface contaminations and partial lattice oxygen from $SiO_2$ surface that creates dangling bonds / active sites due to the formation of localized hot spot on surface of substrates. These localized hot spots are, more energetically favorable to adsorb hydrocarbon species. And the sticking coefficient of hydrocarbon species on the substrate surface is generally temperature dependent. The ECR-CVD creates highly energetic plasma species ($C_xH_y$,$C_2$, CH, $CH^+$,H, $H^+$,etc) through dissociation and recombination of $CH_4$ with Ar.[31] Most effective species are $C_2$, CH, atomic and molecular hydrogen through ECR discharge for the growth of carbon nanostructure.[31] As a result, the rapid nucleation of nanoislands and coalescent between them takes place through direct adsorption and surface diffusion of carbon species on substrate surface.[29] Therefore the migration of energetic plasma species reaching towards the substrate and adsorption rates becomes higher at higher growth temperature.[27] Such substrate temperature dependent growth is also seen in time dependent growth of VGNs. As growth time increases, the continuous NG island transforms into the VGNs.

*Case II* : *Influence of microwave power at constant distance and growth temperature*

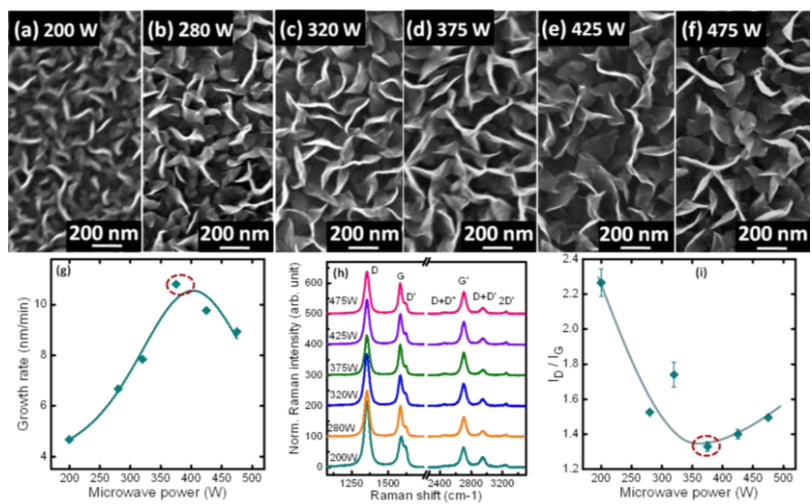

Figure 3 SEM micrographs of VGNs grown at microwave power of (a) 200W, (b) 250W, (c) 320W, (d) 375W, (e) 425W and (f) 475W; (g) variation of growth rate (h) stack of Raman spectra and (i) $I_D/I_G$ of VGNs with respect to plasma power.

The panel (a)-(f) in Fig. 3 reveals that the spatial distribution of vertical sheets and Fig. 3(g) indicates that the growth rates can be controlled by changing the MW power from 200W to 475W, in this study. Interestingly, the higher growth rate is found for the sample grown with 375 W power, while other parameters are fixed. By increasing more power, the growth rates started to decrease as shown in Fig. 3(g). Although height is decreases, the lateral length becomes widening due to continuous availability of hydrocarbon radical as the better width uniformity is observed at high density plasma.[32] Fig. 3(h), represents the stack of Raman spectra of VGNs grown under different MW power, indicates the decrease in D′ intensity with respect to the power. No significant variations are observed in FWHM of D-, G- and G′- band. The Fig. 3(h) shows the variation in $I_D/I_G$ with respect to the MW power, which follows the similar trends as it is observed in growth rate versus MW power plot (Fig. 3(g)). Therefore, in critical equilibrium state between deposition of carbon species onto the substrate and etching of carbon species by hydrogen content, the 375W is found to be optimum plasma power for VGNs growth with highest growth rate and crystalline size, in this study. This result can be explained as follows:

The increase in plasma power, enhance the decomposition rates, density, momentum and temperature of ions, electrons and neutral species. Although the substrates temperature fixed at 800°C for the study, plasma also is sufficient source to heat the substrate temperature locally as well influence on the local electric field intensity.[8, 13] As the MW power increased from 200 to 475 W, the surface adatom mobility increases and electrons move faster towards the substrate. Hence the surface is covered by a *quasi stationary electron film*.[33] This negatively charged surface makes stronger electric field to attract the ion with higher acceleration. It should be noted that the growth of VGNs are mainly based on the competition between deposition rate of carbon species and etching rate of carbon species by nescent H produced in plasma. In addition, the hydrogen species under plasma plays major role during the NG structure growth by PECVD: (i) reducing the amount of $C_2$ species by recombining with them and (ii) etching the $sp^2$-C, $sp^3$-C and a-C at different rates. Hence C/H ratios have a significant impact on shaping the morphology and graphitization which enhance the quality of the structure. Thus initial increase in growth rate and crystalline size with MW power is due to increase in availability in $C_2$ radicals towards the substrate and optimum C/H ratio. At higher power (>375W), the amount and energy of hydrogen species also increases which reduce the $C_2$ radical density as well as etching out carbon species in parallel, resulted in decreased growth rate and crystallinity. So to get optimum graphitic structure, one has to precisely control the amount of hydrogen species by tuning the MW power.

*Case II : Influence of distance between plasma source and substrate at constant microwave power and growth temperature*

Apart from MW power, the distance between substrate to plasma source is another crucial factor to decide the growth kinetics. Fig. 4(a-d) depicts the morphology of VGNs deposited at different distance between substrates and plasma sources by keeping other parameter fixed. It has seen that the number density of vertical sheets decreases, length, height of the sheets and spacing between the sheets of VGNs increases with decreasing the distance. In addition, the growth rates increased from 5 nm/min to 25.5 nm/min as the distance decreased from 40 cm to 10 cm [Fig. 4(e)]. The stack of Raman spectra with respect to the distance [Fig. 4(f)] shows that the D′ intensity for the film grown at various distance. The Fig. 4(g)) confirms that closer the distance, shorter the FWHM of D-, G and G′- peak. The lowest FWHM of G peak is found to 27.9 cm$^{-1}$ for the sample grown at a distance of 10 cm. The $I_D/I_G$ decreases i.e. crystalline size increases with decreasing the distance, as shown in Fig. 4(h). This implies that the growth rates and structural quality of the sample is significantly improved for this sample.

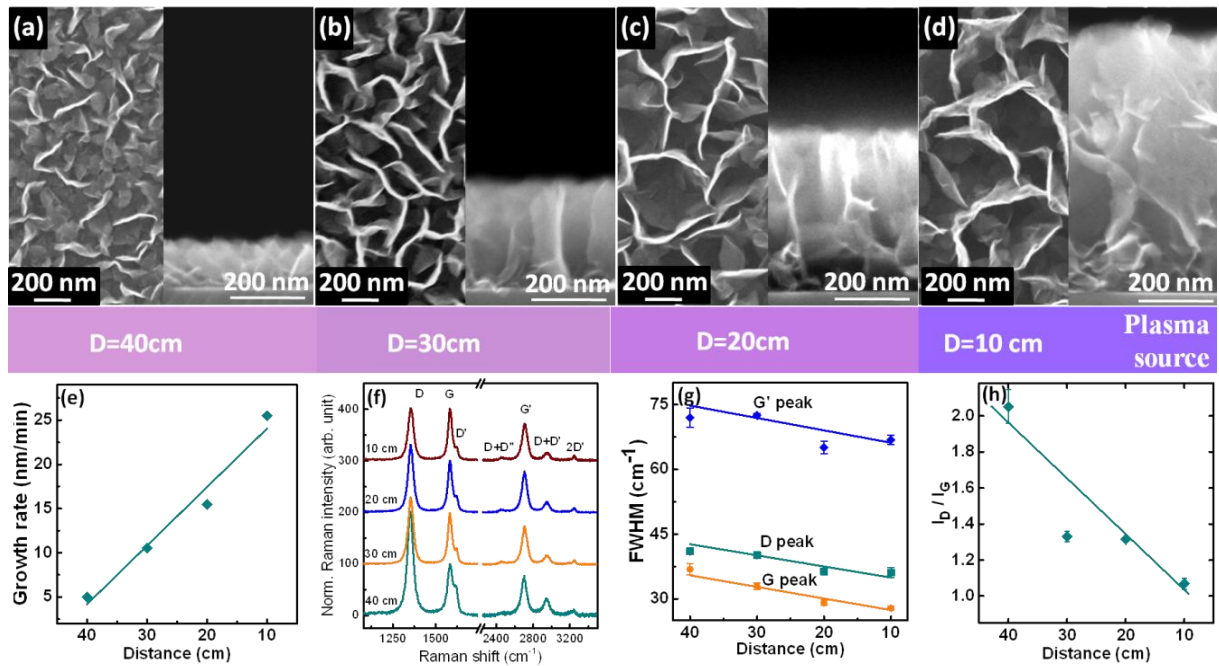

Figure 4.2 SEM micrographs of VGNs grown at a distance from plasma source of (a) 40 cm, (b) 30 cm, (c) 20 cm, and (d) 475W10 cm; (e) variation of growth rate (f) stack of Raman spectra, (g) variation in FWHM of D-, G and G′- peak, (h) variation in $I_D/I_G$ of VGNs with respect to plasma power.

This fact can be explained as follows: at higher distance (40 cm), time of transport of carbon radicals (mainly $C_2$) from plasma source to substrates is high as well they will decay and form other species by recombine with each other. So other way, the radical density and ion energy of plasma species are less at the higher distance. Whereas, at closer to the plasma source, the ion density and ion energy of plasma species are very high and reaching to the substrate with very less amount of recombination. As a result, the growth rate is high for the substrate is placed at 10 cm below the plasma source.

## 4. Conclusion

In summary, vertical graphene nanosheets (VGNs) are synthesized in plasma enhanced chemical vapor deposition by varying substrate temperature, microwave power and distance from substrate to plasma source. The factors contributing here are mainly surface adatom mobility, long-lived species and balance between C/H ratios. In addition to in-built electric field, substrate temperature has observed to be crucial impact for the vertical growth. Hence, it is reflects on morphology and internal structure of the final product. This understanding provides growth of VGNs with optimized and controlled morphology owing to its practical application for field emission, sensor and energy storage.

## Acknowledgement

One of the authors, Subrata Ghosh, acknowledges Department of Atomic Energy, Govt. of India for senior research fellowship. We are thankful to Dr. Sandip Dhara and Dr. A. K. Tyagi for their kind support.